\documentclass[12pt,letterpaper]{article}
\usepackage{amsmath,amssymb,epsfig}
\numberwithin{equation}{section}
\newcommand{\be}{\begin{equation}}
\newcommand{\bea}{\begin{eqnarray}}
\newcommand{\eea}{\end{eqnarray}}
\newcommand{\ba}{\begin{array}}
\newcommand{\ea}{\end{array}}
\newcommand{\ee}{\end{equation}}

\expandafter\ifx\csname mathbbm\endcsname\relax

\else

\fi \textheight 22cm \textwidth 15cm \topmargin 1mm \oddsidemargin
5mm \evensidemargin 5mm

\begin{document}

\begin{titlepage}
 \hfill
 \vbox{
    \halign{#\hfil         \cr
           hep-th/0501072 \cr
           IPM/P-2005/001 \cr
           } 
      }  
 \vspace*{20mm}
 \begin{center}
 {\Large {\bf  Semiclassical String Solutions on 1/2 BPS Geometries }\\ }

 \vspace*{15mm} \vspace*{1mm} {Hajar Ebrahim
 and Amir E. Mosaffa }

 \vspace*{1cm}

 {\it  Institute for Studies in Theoretical
 Physics and Mathematics (IPM)\\
 P.O. Box 19395-5531, Tehran, Iran\\ }

 \vspace*{1cm}
 \end{center}

 \begin{abstract}
 We study semiclassical string solutions on the $1/2$ BPS geometry of type IIB string theory
 characterized by concentric rings on the boundary plane.
We consider both folded rotating strings carrying  nonzero
R-charge and circular pulsating strings. We find that unlike
rotating strings, as far as circular pulsating strings are
concerned, the dynamics remains qualitatively unchanged when the
concentric rings replace $AdS_5\times S^5$. Using the gravity dual
we have also studied the Wilson loop of the corresponding gauge
theory. The result is qualitatively the same as that in
$AdS_5\times S^5$ in the global coordinates where the
corresponding gauge theory is defined on $S^3\times R$. We show
that there is a correction to $1/L$ leading order behavior of the
potential between external objects.

 \end{abstract}

 \end{titlepage}

 \section{Introduction}

The original chain of reasonings leading to AdS/CFT correspondence
relies on considering the near horizon geometry of type IIB D3
brane supergravity solution and conjecturing a relation between
string theory on this geometry and the field theory that lives on
its boundary
\cite{{Maldacena:1997re},{Gubser:1998bc},{Witten:1998qj}}. This
geometry is $AdS_5\times S^5$ in the Poincare coordinates and the
dual theory is $\cal{N}$$=4$ $SU(N)$ SYM on $R^4$. These
coordinates however do not cover all the $AdS$ space and in order
to do so one has to extend to the global coordinates which in turn
amounts to changing the boundary to $S^3\times
R$\cite{Witten:1998qj}. Therefore the global AdS is dual to CFT on
$S^3\times R$.\footnote{We note also that type IIB string theory
on the plane wave limit of the geometry is dual to a quantum
mechanical theory \cite{Sheikh-Jabbari:2004ik}.}

This duality enables us to give geometric (gravitational)
interpretation to different operators in the SYM by identifying
the string excitations that they correspond to. Depending on
energy, these excitations can range from point like field theory
modes to brane configurations which can in
principle cause geometric transition in $AdS_5\times S^5$ due to
back reaction.

Amongst all the operators in $\cal{N}$$=4$ SYM on $S^3\times R$
there is a certain class, known as $1/2$ BPS, which is of special
interest. These preserve half of the original supersymmetry and
are specified by the condition $\Delta-J=0$ where $\Delta$ is the
conformal weight and $J$ is a particular R symmetry charge of the
operator. These operators have a free fermion field theory
description and are characterized by the phase space of the
fermions \cite{Berenstein:2004kk}.

In order to find the geometric counterparts of this class of
operators, the authors of \cite{Lin:2004nb} have established the
general setting for obtaining the corresponding $1/2$ BPS
geometries. This is done by looking for those solutions of type
IIB supergravity equations which have $SO(4)\times SO(4)\times R$
isometry and which solve the killing spinor equations. Doing so,
one picks from all the excitations in $AdS_5\times S^5$ in the
global coordinates, those which constitute its $1/2$ BPS sector.
It turns out that these symmetry requirements, plus regularity,
are very restrictive such that the whole solution is determined by
a single function which should satisfy a linear differential
equation subject to certain boundary conditions on a 2-plane. The
phase space of the underlying free fermion system is then
identified with the different allowed configurations for this
function on the 2-plane.

These families of solutions for constant axion and dilaton and
zero three-form field strengths are given by
 \bea
 ds^2&=&-h^{-2}(dt^2+V_idx^i)^2+h^2(dy^2+dx^idx^i)+ye^Gd\Omega_3^2
 +ye^{-G}d{\tilde\Omega}_3^2\;,\cr && \cr
 h^{-2}&=&2y\cosh{G}\;,\cr && \cr
 y\partial_yV_i&=&\epsilon_{ij}\partial_jz\;,\;\;\;\;\;
 y(\partial_iV_j-\partial_jV_i)=\epsilon_{ij}\partial_yz\;,\;\;\;\;\;
 z=\frac{1}{2}\tanh{G}\;,
 \label{bub}
 \eea
where $i,j=1,2$ and $z$ satisfies the following equation
 \be
 \partial_i\partial_jz+y\partial_y(\frac{\partial_yz}{y})=0\;.
 \ee
To get a nonsingular solution, $z$ must obey the boundary
condition $z=\pm1/2$ at $y=0$ on the 2-plane $(x_1,x_2)$. The self
dual five-form field strength is also nonzero (for details see
\cite{Lin:2004nb}). One can now start with different allowed
boundary conditions for $z$ and obtain the corresponding
solutions. For further studies in this direction see
\cite{Buchel:2004mc}-\cite{Filev:2004yv}.

One would naturally like to study string excitations on each of
the above backgrounds. String modes in different parts of the
background can be studied by expanding string sigma model around
classical configurations. These represent strings that are
propagating in different parts of the space. The modes can in
principle be non BPS and the deviation from BPS condition can be
tuned by changing the classical charges of the configuration such
as spin, angular momentum and etc. An important lesson from the
semiclassical analysis of strings (see for example 
\cite{Gubser:2002tv,Frolov:2002av,semi} and also 
\cite{Tseytlin:2004xa} for reviews) is that when the charges are very large, a
lot can be learned from the classical limit itself. This becomes
an even better approximation when the number of charges is
increased.

In a recent paper \cite{Filev:2004yv}, rotating folded strings
(first studied in \cite{Gubser:2002tv}) have been considered in
the $1/2$ BPS geometry of type IIB which is characterized by the
concentring rings configuration on the $(x_1,x_2)$ plane. This
background is time independent and in certain limits can be
thought of as a configuration of smeared $S^5$ giants and/or their
$AdS_5$ duals. The corresponding metric in the polar coordinates
is given by
 \bea
 ds^2&=&-h^{-2}(dt+V_rdr+V_{\phi}d\phi)^2+h^2(dy^2+dr^2+r^2d\phi^2)+
 ye^{G}d\Omega_3^2+ye^{-G}d{\tilde\Omega}_3^2\;,
 \cr &&\cr
 h^{-2}&=&2y\cosh(G),\;\;\;\;\;\;\;\;\;\;\;\;e^{G}=\sqrt{\frac{1+{\tilde z}}{ -{\tilde
 z}}}\;,
 \label{bac}
 \eea
where
 \bea
 {\tilde z}&=&\frac{1}{2}\sum_{n=1}^N    (-1)^{n+1}\left(\frac{r^2-r^2_n+y^2}
 {\sqrt{(r^2+r_n^2+y^2)^2-4r^2r_n^2}}-1\right)\;,
 \cr &&\cr V_r&=&0\;,\cr &&\cr
 V_\phi&=&\frac{1}{2}\sum_{n=1}^N   (-1)^n\left(\frac{r^2+r^2_n+y^2}
 {\sqrt{(r^2+r_n^2+y^2)^2-4r^2r_n^2}}-1\right)\;.
 \eea
Here $r_1$ is the radius of the outermost circle, $r_2$ the next
and so on. In the case of one radius, this is just one
$AdS_5\times S^5$. We will only consider the case where $N$ is an
odd number, therefore we will have a black disc in the middle of
the configuration and $N-1$ rings. To fix our notation we may
parameterize the two spheres as follows
 \bea
 d\Omega_3^2&=&d\theta^2_1+\cos^2\theta_1(d\theta_2^2+\cos^2\theta_2d\theta^2)\;,\cr &&\cr
 d{\tilde
 \Omega}_3^2&=&d\psi_1^2+\cos^2\psi_1(d\psi_2^2+\cos^2\psi_2d\psi^2)\;.
 \eea

It was found in \cite{Filev:2004yv} that as compared to
$AdS_5\times S^5$, the concentric rings provide new physics for
rotating strings. That is, a folded rotating string can have
orbital angular momentum in $S^3$ in addition to the spin about
its center of mass. Such orbiting strings had already been found
in confining $AdS$ backgrounds \cite{Armoni:2002xp,Armoni:2002fr}
and in this sense the concentric rings show some sort of
similarities with such backgrounds.

In the present work we continue the semiclassical analysis of
strings in concentric rings. We start by generalizing the folded
string of \cite{Filev:2004yv} to the case with nonzero angular
momentum in $\tilde{S}^3$ and then focus on circular pulsating
strings. These string configurations were first studied in
$AdS_5\times S^5$ in \cite{Minahan:2002rc} and were then
generalized to different situations \cite{pulsating}. We consider
the concentric rings as a deformation of $AdS_5\times S^5$ by
taking the outermost radius to be much larger than the rest. This
can be thought of as a configuration of a number of giants which
are located near the pole of $S^5$ and close to one another and
smeared in the polar coordinate of $S^5$.

We use Bohr-Sommerfeld analysis to find the energy levels of the
pulsating string in terms of the level quantum number when the
energy is very large. Our results show that unlike the rotating
strings, pulsating ones experience no new physics and the dynamics
is qualitatively that in the $AdS_5\times S^5$ background. The
rings affect the dynamics slightly only when the radius of the
string becomes comparable to their radii.

The paper is organized as follows. In section 2, we will study
rotating and folded closed strings in both long and short string
limits on 1/2 BPS geometry (\ref{bac}) where the relation between
their energy and spin is obtained. We will also consider another
background which could be found by taking the Penrose limit of the
concentric rings configuration. In section 3, we shall study the
circular pulsating strings on this background that its energy
relation is concluded. In section 4 we will study Wilson loop of
the corresponding dual theory using open strings on this
background. The last section is devoted to conclusions.


\section{Rotating and Spinning folded strings}

In this section we shall study semiclassical closed strings in the
1/2 BPS geometry (\ref{bac}) which is extended in the $y$
direction while rotating in both 3-spheres. Using our notation the
corresponding closed string configuration is given by
 \be
 t=\kappa \tau,\;\;\;\;\theta=\omega \tau,\;\;\;\;\psi=\nu \tau,\;\;\;\;y=y(\sigma)\;,
 \label{ans}
 \ee
and all other coordinates are set to zero. This closed string
configuration has recently been studied in \cite{Filev:2004yv}
where the authors have found the dependence of the energy of the
string on the spin which represents the string's rotation in the
first sphere of the solution (\ref{bac}). Here we will study a
more general case where the string has nonzero angular momentum in
both spheres.

The bosonic part of the GS superstring action
for this closed string configuration on the
background (\ref{bac}) is
 \be
 S=\frac{1}{4\pi}\int d^2\sigma \;\left(h^{-2}\kappa^2+h^2{y'}^2-\omega^2
 ye^{G}-\nu^2 ye^{-G}\right)\;.
 \ee

For generic values of $\omega$ and $\nu$ this ansatz describes a
classical string which is stretched and folded along $y$ and
rotates in the $\theta$ and $\psi$ directions. The corresponding
conserved charges of the system are
 \be
 E=\frac{\kappa}{2\pi}\int_0^{2\pi}d\sigma \;2y\cosh(G),\;\;\;\;\;
 S=\frac{\omega}{2\pi}\int_0^{2\pi} d\sigma\;ye^{G},\;\;\;\;\;
 J=\frac{\nu}{2\pi}\int_0^{2\pi}d\sigma\; ye^{-G}\;.
 \ee
Now the aim is to find the dependence of energy $E$ on $S$ and
$J$. To find this one may use the equation of motion derived from
this action. We note however that it is useful to work with the
first integral of the equation of motion which in this case is the
Virasoro constraint
 \be
 {y'}^2=y^2\bigg{[}(\omega^2-\kappa^2)(1+e^{2G})-(\kappa^2-\nu^2)(1+e^{-2G})\bigg{]}\;.
 \ee
Using the expression for $G$ we arrive at
 \be
 {y'}^2=y^2(\kappa^2-\nu^2)\bigg{(}\frac{1}{1+{\tilde z}}+\eta\;
 \frac{1}{{\tilde z}}\bigg{)}\;,
 \ee
where $\eta=\frac{\omega^2-\kappa^2}{\kappa^2-\nu^2}$. In terms of
this parameter the turning points of the string along the $y$
direction is given by $\frac{{\tilde z}_0}{1+{\tilde z}_0}=\eta$.
Therefore in the simplest one folded case the interval $0\leq
\sigma\leq 2\pi$ is split into 4 segments in which for $0\leq
\sigma \leq \frac{\pi}{2}$ the function $y(\sigma)$ increases from
zero to its maximal value given by $z_0$.

Using the definition of the energy, $E$, spin $S$ and the angular
momentum $J$ of the string one sees that
 \be
 E=\frac{\kappa}{\nu}\;J+\frac{\kappa}{\omega}\;S \label{rel}\;,
 \ee
which together with the Virasoro constraint could be used to
determine the dependence of $E$ on $S$ and $J$. Following
\cite{Gubser:2002tv} we shall study the limits of short string
($\eta\rightarrow \infty$) and long string ($\eta\rightarrow 0$),
separately.


\subsection*{Short strings}
The short string limit corresponds to the case where $\eta
\rightarrow \infty$. In this case one may expand ${\tilde z}$
around $y = 0$
 \be {\tilde
 z}=\sum_{n=1}^N(-1)^n\frac{r_n^2}{r_n^2+y^2}\sim
 -1+f_0y^2-f_1y^4\;,
 \ee
where
 \be
 f_0=\sum_{n=1}^N(-1)^{n+1}\;\frac{1}{r_n^2},\;\;\;\;\;f_1=\sum_{n=1}^N(-1)^{n+1}\;
 \frac{1}{r_n^4}\;. \ee By making use of this expression one finds \be
 {y'}^2=\frac{\kappa^2-\nu^2}{f_0}-(\kappa^2-\nu^2)\left(\eta-\frac{f_1}{f_0^2}\right)y^2
 +{\cal O}(y^4)\;.
 \ee
The condition for having a singly folded string with radius
$y_0\ll 1$ is
 \be
 (\kappa^2-\nu^2)\left(\eta-\frac{f_1}{f_0^2}\right)=1,\;\;\;\;\;y_0^2=\frac{\kappa^2-\nu^2}
 {f_0}\;,
\ee and therefore in leading order we get
 \be
 \kappa^2-\nu^2\sim
 \frac{1}{\eta},\;\;\;\;\;\omega^2=\nu^2+1+\frac{1}{\eta}\;.
 \ee

On the other hand we have
 \be
 ye^{G}\sim y^2\sqrt{f_0}\bigg{[}1+\frac{f_0}{2}\left(1-\frac{f_1}{f_0^2}\right)
 y^2\bigg{]},\;\;\;\;
 ye^{-G}\sim \frac{1}{\sqrt{f_0}}\bigg{[}1-\frac{f_0}{2}\left(1-\frac{f_1}{f_0^2}\right)
 y^2\bigg{]},
 \ee
which can be used to obtain the spin and angular momentum up to
$\mathcal{O}(y^n)$ as follows
 \bea
 S &\approx&\frac{\omega}{2\pi}\int_0^{2\pi}d\sigma\;
 \frac{\kappa^2-\nu^2}{\sqrt{f_0}}\;\sin^2\sigma\bigg{[}1+\frac{\kappa^2-\nu^2}{2}
 \left(1-\frac{f_1}{f_0^2}\right)
 \;\sin^2\sigma\bigg{]}\;,\cr &&\cr &&\\
 J&\approx&\frac{\nu}{2\pi}\int_0^{2\pi}d\sigma\;
 \frac{1}{\sqrt{f_0}}\;\bigg{[}1-\frac{\kappa^2-\nu^2}{2}\left(1-\frac{f_1}{f_0^2}\right)
 \;\sin^2\sigma\bigg{]}\;.\nonumber
 \eea
Here we have used the fact that
$y^2=\frac{\kappa^2-\nu^2}{f_0}\sin^2\sigma$. By making use of
$\kappa^2-\nu^2\sim \frac{1}{\eta}$ one finds
 \bea
 S&\sim &
 \frac{\omega}{2\sqrt{f_0}}\;
 \frac{1}{\eta}\bigg{[}1+\frac{3}{8}\left(1-\frac{f_1}{f_0^2}\right)\;\frac{1}{\eta}
 \bigg{]}\;,\cr &&\cr &&\cr
 J&\sim&\frac{\nu}{\sqrt{f_0}}\bigg{[}1-\frac{1}{4}
 \left(1-\frac{f_1}{f_0^2}\right)\;\frac{1}{\eta}
 \bigg{]}\;.
 \eea
Altogether in leading order we arrive at
 \be
 \nu=\sqrt{f_0}J,\;\;\;\;\;
 \kappa^2\sim f_0J^2+\frac{2\sqrt{f_0}S}{\sqrt{f_0J^2+1}},\;\;\;\;\;
 \omega^2\sim 1+f_0J^2+\frac{2\sqrt{f_0}S}{\sqrt{f_0J^2+1}}.
 \ee
Plugging these into (\ref{rel}) one gets
 \be
 E\approx\sqrt{J^2+\frac{2S}{\sqrt{f_0^2J^2+f_0}}}\;
 \bigg{(}1+
 \frac{S}{\sqrt{J^2+\frac{2S}{\sqrt{f_0^2J^2+f_0}}+\frac{1}{f_0}}}
 \bigg{)}\;.
 \ee

For the situation where both $J$ and $S$ are small one finds
 \be
 E^2\approx J^2+\frac{2S}{\sqrt{f_0}}\;,
 \ee
which actually represents the Reggae trajectories in the flat space. On the other hand
for the case where  $J\ll S$ we get
 \be
 E\approx
 \frac{\sqrt{2S}}{f_0^{1/4}}+\frac{J^2}{2f_0^{1/4}\sqrt{2S}}\;,
 \ee
which, for the limit of $J\gg 2S$, it leads to
 \be
 E\approx J+S+\frac{S}{2f_0J^2}\;.
 \ee

We note that these are exactly the same expressions found in
\cite{Frolov:2002av}, where the authors have studied the same string
configuration as (\ref{ans}) on the $AdS_5\times S^5$, if we
identify $\lambda$ with $1/f_0$ where $\lambda$ is radius of the
AdS space.


\subsection*{Long strings}

In the long string limit we have $\eta\rightarrow 0$ where the
maximal value of $y_0$ is large. In this case one may expand
${\tilde z}$ for large $y$
 \be
 z=\sum_{n=1}^N(-1)^n\frac{r_n^2}{r_n^2+y^2}\approx
 -\frac{g_0}{y^2}+\frac{g_1}{y^4}\;,
 \ee
where
 \be
 g_0=\sum_{n=1}^N(-1)^{n+1}r_n^2,\;\;\;\;\;\;
 g_1=\sum_{n=1}^N(-1)^{n+1}r_n^4\;,
 \ee
therefore one finds
 \be
 {y'}^2=y^2(\kappa^2-\nu^2)(1-\frac{\eta}{g_0}\;y^2)\;.
 \label{longy}
 \ee
We note that for $0<\sigma <\frac{\pi}{2}$ the function
$y(\sigma)$ increases from zero to its maximal value $y_0$ given
by $y_0=\sqrt{g_0/\eta}$, so
 \be
 2\pi=\int_0^{2\pi}d\sigma=\frac{4}{\sqrt{\kappa^2-\nu^2}}
 \int_0^{y_0}\frac{dy}{y\sqrt{1-\frac{\eta}{g_0}\;y^2}}\;.
 \ee
Therefore we get
 \be
 \kappa^2\sim \nu^2+\frac{1}{\pi^2}\ln^2\frac{g_0}{\eta},\;\;\;\;\;\;\;
 \omega^2\sim \nu^2+ \frac{1}{\pi^2}(1+\eta)
 \ln^2\frac{g_0}{\eta}\;.
 \ee
On the other hand we find
 \bea
 ye^{G}&\sim&\frac{y^2}{\sqrt{g_0}}\left(1-\frac{g_0}{2}(1-\frac{g_1}{g_0^2})\;
 \frac{1}{y^2}\right)\;,\cr &&\cr
 ye^{-G}&\sim&\sqrt{g_0}\left(1+\frac{g_0}{2}(1-\frac{g_1}{g_0^2})\;
 \frac{1}{y^2}\right)\;.
 \eea
Plugging these in the expressions of $S$ and $J$ and using the equation
(\ref{longy}) we arrive at
 \be
 S\approx \frac{\omega\sqrt{g_0}}{\eta\ln\frac{g_0}{\eta}},
 \;\;\;\;\;J=\sqrt{g_0}\nu\;.
 \ee
For the case of $\nu\ll \ln\frac{g_0}{\eta}$ this results
$\frac{\sqrt{g_0}}{\eta}\sim \frac{\pi S}{2}$
and by making use of
the relation (\ref{rel}) we find
 \be
 E\approx S+\frac{\sqrt{g_0}}{\pi}\ln\frac{S}{\sqrt{g_0}}+
 \frac{\pi J^2}{2\sqrt{g_0}\ln\frac{S}{\sqrt{g_0}}}\;.
 \ee
On the other hand in the opposite limit where $\nu\gg \ln\frac{g_0}{\eta}$  we get
 \be
 E\approx S+J+\frac{\sqrt{g_0}}{2\pi^2 J}\ln^2\frac{S}{J}\;.
 \ee
We note that in comparison with the AdS case studied in \cite{Frolov:2002av}
we get the same expressions if we identify
the AdS radius with $\sqrt{g_0}$.

So far it seems that the folded closed strings (\ref{ans}) on the
1/2 BPS geometry (\ref{bac}) qualitatively feel the same physics
as the AdS background as far as the large and small energy regimes
are concerned. We note however that in the intermediate scale the
new physics might appear. In fact it was shown \cite{Filev:2004yv}
that this is the case at least for the rotating folded closed
strings. This can be understood as follows.

From the first integral of motion we see that the condition for
having a folded string is equivalent to the condition for having
two turning points for a one-dimensional motion in the effective
potential \cite{Filev:2004yv}
 \be V(y)=-\frac{1}{{\tilde
 z}}=\frac{-1}{\sum (1)^{n+1} \frac{r_n^2}{r_n^2+y^2}}\;.
 \ee
This potential has several minima and therefore going from large
scale into the intermediate scale the string can  split into
smaller folded orbiting strings which are located around these new
minima.

There is an other interesting situation one may have because of
these new minima. In fact one can consider a folded closed string
localized around one of the minima with large $J$ charge
corresponding to the high angular momentum in the $\psi$
direction. In this situation, if we had considered AdS case, we
would have got the plane wave background for the quadratic
fluctuations around this classical solution. In this case we would
also expect to get the plane wave solution for each minima. One
might also suspect that cutting a small strip, say around
$\phi=\pi/2$, in the boundary plane of the solution (\ref{bac}),
could result in a new solution which is the superposition of the
plane wave solutions we get from each minimum.

To be more precise, we consider the concentric rings background as
a deformation of $AdS_5\times S^5$ by demanding that $r_n-r_N\ll
r_N$. This can be viewed as an $AdS_5\times S^5$ with the radius
$r_1 (r_N)$ containing a number of giants (AdS giants) which are
located close to one another and close to the equator of $S^5$,
$\theta=0$, smeared in the polar coordinate. With this assumption
we expand the rings around the point $r=r_N$, $y=0$ and
$\phi=\pi/2$ by defining
 \be
 r-r_N\equiv\frac{x_2}{r_N}\;,\;\;\;\;\;y\equiv\frac{w}{r_N}\;,
 \;\;\;\;\;r_n-r_N\equiv\frac{R_n}{r_N}\;,\;\;\;\;\;
 \phi-\frac{\pi}{2}\equiv-\frac{x_1}{r_N^2}\;.
 \ee
Writing the concentric rings solution in terms of the above
parameters, taking the limit of $r_N\rightarrow\infty$ and
renaming $w$ as $y$, will result in a background which is the
superposition of a number of plane waves
 \bea
 z(x_2,y)&=&\frac{1}{2}\sum_{n=1}^N(-1)^{n+1}\frac{x_2-R_n}{\sqrt{(x_2-r_n)^2+y^2}}
 \;\;\;\;\;\;(R_N=0)\;,
 \cr
 &&\cr
 V_1&=&V_{\phi}\;\partial_{x_1}\phi\;=\frac{1}{2}\sum_{n=1}^N(-1)^{n+1}
 \frac{1}{\sqrt{(x_2-R_n)^2+y^2}}\;,\;\;\;V_2=0\;.
 \label{zebra}
 \eea

This background can be found directly from the equations of motion
by demanding the following boundary condition on the $(x_1,x_2)$
plane \cite{Lin:2004nb}
 \be
 z(x'_1,x'_2,0)=\frac{1}{2}\sum_{n=1}^{N}
 (-1)^{n+1}Sign(x'_2-R_n)\;,
 \ee
where we take $N$ to be an odd number such that the solution
becomes pp wave asymptotically. This boundary condition is shown
by a number of horizontal black and white strips on the
$(x_1,x_2)$ plane bounded by black and white regions from below
and above respectively and thus we call its corresponding solution
the "{\it Zebra}" background. It is easy to show that this
boundary condition will result in the solution (\ref{zebra}).



\section{Circular Pulsating Strings }
We now study pulsating strings in the background (\ref{bac}).
These string solutions were first studied in \cite{Minahan:2002rc} on
$AdS_5\times S^5$ and generalizations to other backgrounds were
given in \cite{pulsating}. Let us first briefly review the $AdS_5\times
S^5$ case. A pulsating string is defined through the following
ansatz for the embedding coordinates
 \be
 t=\tau\;,\;\;\;\;\;\rho=\rho(\tau)\;,\;\;\;\;\;\theta=m\sigma\;,
 \ee
where $\rho$ is the radial direction of $AdS$ in the global
coordinates and $\theta$ is a great circle of the $S^3$ contained
in $AdS_5$. The rest of the coordinates are taken to be zero.
Writing the Nambu-Goto action for this configuration, we arrive at
a one dimensional quantum mechanical system with the Hamiltonian
 \be
 H=\sqrt{\Pi^2+m^2\lambda\tan^2\xi\sec^2\xi}\;,
 \ee
where $\lambda$ is the 't Hooft coupling,
$\xi=\sin^{-1}(\tanh\rho)$ and $\Pi$ is the conjugate momentum of
$\xi$. One can then define a potential
 \be
 V(\xi)=m^2\frac{\lambda\tan^2\xi}{\cos^2\xi}
 \label{pot1}\;,
 \ee
for the system with the Hamiltonian $H^2$ and find the energy
levels.

With this brief review we now turn to the problem of a pulsating
string in (\ref{bac}). We consider the following ansatz
 \be
 t=\tau\;,\;\;\;\;\;y=y(\tau)\;,\;\;\;\;\;\theta=m\sigma\;,
 \ee
and the rest of coordinates are zero. The Nambu-Goto action for
this configuration reads
 \be
 S_{NG}=-\frac{m}{\alpha'}\int dt\; g(\xi)\sqrt{1-\dot{\xi}^2}\;,
 \ee
where
 \be
 \frac{d\xi}{dy}=h^2\;,\;\;\;\;\;g(\xi)=\frac{y}{\sqrt{-\tilde{z}}}\;.
 \ee
We find the momentum and Hamiltonian for the system as
 \be
 \Pi=\frac{m}{\alpha'}\;g(\xi)\;\frac{\dot{\xi}}{\sqrt{1-\dot{\xi}^2}}\;,
 \;\;\;\;\;H=\sqrt{\Pi^2+\bigg(\frac{m}{\alpha'}\bigg)^2\;g(\xi)^2}\;.
 \ee
Now one can consider $H^2$ as a one dimensional system with the
potential
 \be
 V(\xi)=\bigg(\frac{m}{\alpha'}\bigg)^2\;g(\xi)^2\;.
 \label{pot2}
 \ee
To compare this potential with the one in (\ref{pot1}), we define
the variables $\xi_n$ and the constants $\lambda_n$ by
 \be
 \cos^2\xi_n=\frac{r_n^2}{y^2+r_n^2}\;,\;\;\;\;\;\sin^2\xi_n=\frac{y^2}{y^2+r_n^2}\;,
 \;\;\;\;\;\lambda_n=\frac{r_n^2}{\alpha'^2}\;,
 \ee
where it is clear that these variables are not independent, as
they are all determined by $y$, and vary between zero and
$\xi_n^{max}$ which are determined by $y_{max}$. One should also
note that $\xi_1<\xi_2<\cdots<\xi_N$ for a given value of $y$. In
terms of these parameters the potential can be brought to a form
which can be easily compared with (\ref{pot1})
 \be
 V=\frac{m^2}{N}\frac{\sum_{n=1}^N\lambda_n\tan^2\xi_n}{\sum_{n=1}^N\cos^2\xi_n}\;.
 \ee
For $N=1$, the above potential reduces to (\ref{pot1}). The
qualitative behavior of this potential is the same as that in the
$AdS_5\times S^5$ case; classically the circular string pulsates
between a point and a circle whose radius is determined by the
string's energy. So, as far as pulsating strings are concerned, no
new physics emerges in this problem. This should be compared with
the folded strings of the previous section for which the new
exterma in the potential give rise to orbiting strings in addition
to spinning ones.

We can now proceed to find the energy levels of our one
dimensional quantum mechanical system with the potential given by
(\ref{pot2}). We are mainly interested in string states with a
large quantum number which, in our problem, is the level number.
Therefore we focus on large energy states for which the
Bohr-Sommerfeld analysis is a good approximation. As the energy of
the string increases, the $\xi_n^{max}$ get closer to $\pi/2$.
Since $y$ is a radial coordinate we will symmetrize the potential
around $y=0$ by allowing the $\xi_n$ to range between $-\pi/2$ and
$\pi/2$ and consider only the even wave functions
 \be
 (2n+1/2)\pi\approx\int_{-\xi_0}^{\xi_0}
 d\xi\sqrt{E^2-\bigg(\frac{m}{\alpha'}\bigg)^2\;g(\xi)^2}\;,\;\;\;\;\;
 n=0,1,2,\cdots\;,
 \label{bs}
 \ee
where $\pm\xi_0$ are found from the equation
 \be
 E=\bigg(\frac{m}{\alpha'}\bigg)\;g(\xi_0)\;.
 \label{eta}
 \ee
To perform the integrations we consider the background (\ref{bac})
as a deformation of $AdS_5\times S^5$ by assuming that the radius
of the outermost circle is much larger than the rest
 \be
 R\equiv
 r_1\;,\;\;\;\;\;\beta_n\equiv\frac{r_n}{R}\ll1\;\;\;(1<n\leq N)\;.
 \ee
We define the large energy limit of the string by $B\equiv
E(m\sqrt{\lambda_1})^{-1}\gg1$. The limit of large $B$ and small
$\beta$ means that the string spends most of its time in the
region where $\xi_n\;(1<n\leq N)$ have almost saturated their
bound at $\pi/2$ and therefore it is as if the string is pulsating
in $AdS_5\times S^5$. The new features of the background
(\ref{bac}) play a part in the dynamics only when the string has
shrunk to a small circle. The radius of this circle becomes
arbitrarily small as $\beta\rightarrow0$ and the maximum radius of
the pulsating string becomes arbitrarily large as
$B\rightarrow\infty$. With these assumptions we can write the
following expansions for $h$ and $V$ around their values in the
$AdS_5\times S^5$ background
 \bea
 h^2&\approx&\frac{1}{BR}\;\frac{1}{B^{-1}+x^2}\;\bigg[1+\frac{1}{2}\;
 \beta^2\;\frac{B^{-2}-x^4}{x^4}\bigg]\;,\cr && \cr
 V&\approx& E^2x^2(B^{-1}+x^2)\;\bigg[1+\beta^2\frac{B^{-1}+x^2}{x^2}\bigg]\;,
 \eea
where we have defined
 \be
 x=\frac{y}{R\sqrt{B}}\;,\;\;\;\;\;\;\;
 \beta^2=\sum_{n=2}^{N}(-1)^n\beta_n^2=1-\frac{g_0}{R^2}\;.
 \ee
It is useful to write the integral in (\ref{bs}) as the sum of two
integrals as following
 \bea
 \frac{2mR}{\alpha'}\sqrt{B}&\bigg\{&\int_0^{x_0}\frac{dx}{B^{-1}+x^2}\;
 \bigg[1+\frac{1}{2}\;
 \beta^2\;\frac{B^{-2}-x^4}{x^4}\bigg]\cr && \cr
 &-&\int_0^{x_0}\frac{dx}{B^{-1}+x^2}\;\bigg[1+\frac{1}{2}\;
 \beta^2\;\frac{B^{-2}-x^4}{x^4}\bigg][1-\sqrt{1-V/E^2}]\bigg\}\;.
 \label{int}
 \eea
The first integral is nothing but $\xi_0$ which can be found as
 \be
 \xi_0=BR\int_0^{x_0}dx\;h^2\approx\tan^{-1}(\sqrt{B}x_0)+\frac{1}{2}\beta^2\;
 \frac{3Bx_0^2-1}{3B^{3/2}x_0^3}\;.
 \ee
Noting that $x=1/\sqrt{B}\tan\xi_1$, we find from the above
relation that $\xi_0$ is approximately $\xi_1^{max}$ (for small
$\beta$) which is very close to $\pi/2$ (for large $B$). One can
use (\ref{eta}) to find
 \be
 \xi_0\approx\frac{\pi}{2}-\frac{1}{\sqrt{B}}\;(1-\frac{1}{4}\;\beta^2)\;.
 \ee

The second integral in (\ref{int}) remains finite in the large $B$
limit and is written as
 \be
 (1-\frac{1}{4}\;\beta^2)\int_0^{1}\frac{du}{u^2}(1-\sqrt{1-u^4})
 =(1-\frac{1}{4}\;\beta^2)\bigg(-1+\frac{(2\pi)^{3/2}}{\Gamma(\frac{1}{4})^2}\bigg)\;.
 \ee
Therefore the large energy approximation takes the following form
 \be
 (2n+1/2)\pi\approx
 E\pi-(1-\frac{1}{4}\;\beta^2)\frac{4\pi(2\pi)^{1/2}}{\Gamma(\frac{1}{4})^2}
 m^{1/2}\lambda_N^{1/4}\sqrt{E}\;.
 \ee
We can invert this relation and find
 \be
 E\approx2n+(1-\frac{1}{4}\;\beta^2)\frac{8\pi^{1/2}}{\Gamma(\frac{1}{4})^2}
 \lambda_N^{1/4}\sqrt{mn}\;.
 \ee
This is our final expression for the large energy states of a
circular string, in terms of a large number $n$, when it pulsates
in the background (\ref{bac}) which is considered as a deformation
of $AdS_5\times S^5$.

\section{Wilson loop}

In the previous sections we have studied semiclassical closed
strings in the 1/2 BPS geometries. We note however that the open
string configurations in a given gravity background could also be
used to study the Wilson loop and thereby the potential between
external objects in the corresponding dual field theory
\cite{{Maldacena:1998im},{Rey:1998ik}}.

In this section we shall study the open string solutions in the
1/2 BPS geometries. One may then identify this with the Wilosn
loop in the dual gauge theory which is presumably living on the
boundary with the topology of $S^3\times R$.

To warm up we will first study the Wilson loop in ${\cal N}=4$ SYM
theory on $S^3\times R$. This can be done using the corresponding
supergravity description which in this case is $AdS_5\times S^5$
in the global coordinates given by
 \bea
 ds^2&=&-R^2(1+\frac{r^2}{R^2})dt^2+\frac{dr^2}{1+\frac{r^2}{R^2}}+r^2d\Omega_3^2+R^2
 d\Omega_5^2\;,\cr &&\cr
 d\Omega_3^2&=&d\theta_1^2+\cos^2\theta_1(d\theta_2^2+\cos^2\theta_2\;d\theta^2)\;.
 \eea
Here we have used a unit in which $t$ is dimensionless.

Let us now consider the following open string configuration
 \be
 t=\tau,\;\;\;\;\;\theta=\sigma,\;\;\;\;\;\;r=r(\sigma),
 \;\;\;\;\;\;\theta_1=\theta_2=0\,.
 \ee
The classical string action for the above string configuration is obtained as the
following
 \be
 S=\frac{R}{2\pi\alpha'}\int
 dt\;d\theta\;\sqrt{{r'}^2+r^2(1+\frac{r^2}{R^2})}\;,
 \ee
where prime denotes derivative with respect to $\theta$. Since the
action is $\theta$ independent, the corresponding Hamiltonian is
constant of motion leading to
 \be
 \frac{\frac{r^4}{R^2}(1+\frac{R^2}{r^2})}{\sqrt{{r'}^2
 +\frac{r^4}{R^2}(1+\frac{R^2}{r^2})}}
 =\frac{r_0^2}{R}\sqrt{1+\frac{R^2}{r_0^2}}\;,
 \ee
where $r_0$ is the point where $r'=0$.

Setting $y=\frac{r}{r_0}$ and $\epsilon=\frac{R}{r_0}$ one may
find the distance between two external objects in the theory as
follows
 \be
 \frac{\theta}{2}=\frac{R}{r_0}\int_1^\infty\frac{dy}{y^2(1+\epsilon^2/y^2)^{1/2}
 \sqrt{y^4\frac{1+\epsilon^2/y^2}{1+\epsilon^2}-1}}\;.
 \ee
The potential energy is given by
 \be
 E=\frac{Rr_0}{2\pi\alpha'}\left[\int_1^\infty
 dy\left(\frac{y^2}{\sqrt{y^2-\frac{1+\epsilon^2}{1+\epsilon^2/y^2}}}-1\right)-1\right]\;.
 \ee

The aim is now to eliminate $r_0$ between these expression to find
$E$ in terms of $\theta$. In general it is difficult to do so.
Nonetheless one may work in the limit where $\epsilon \ll 1$. In
the limit of $\epsilon\rightarrow 0$ one would expect to get the
same result as in the theory on the Minkowski space which is dual
to the theory on AdS in Poincare patch. For small $\epsilon$ one
then expects to get some corrections which could be due to short
distance effects taking into account that in this case the theory
is defined on a sphere.

In fact expanding the above expression in $\epsilon$ one gets
 \bea
 \frac{\theta}{2}&=&\frac{R}{r_0}\int_1^\infty \frac{dy}{y^2\sqrt{y^4-1}}
 \left(1+\frac{y^4-y^2-1}{2y^2(1+y^2)}\;
 \epsilon^2+\cdots\right),\cr &&\cr
 E&=&\frac{Rr_0}{2\pi \alpha'}\int_1^\infty dy\left[\frac{y^2}{\sqrt{y^4-1}}
 \left(1+\frac{\epsilon^2}
 {2y^2(1+y^2)}-\frac{(4y^2+1)\epsilon^4}{8y^4(1+y^2)^2}+\cdots\right)-1\right]\cr &&\cr
 &-&\frac{Rr_0}{2\pi \alpha'}\;.
 \eea
Therefore in leading order we find
 \be
 E\sim -\frac{R^2}{2\pi\alpha'}\;\frac{1}{\theta}\;(1+c_0
 \theta^2+c_1\theta^2\cdots)\;,
 \ee
where $c_i$'s are some numerical constants. As we see, besides the
standard $R^2/\theta$ term we have some corrections which could be
understood due to short distance effects. We note that corrections
to the Wilson loop, we have here, might be related to those in
\cite{cusp} where the authors have studied the  Wilson loop in
gauge theory side when there is a cusp in the loop. There the
authors have found corrections to the Wilson loop proportional to
the cusp angle. One might then wonder that our corrections have
the same origin under mapping to plane.\footnote{We would like to
thank Albion Lawrence for a discussion on this point.}

Let us now consider the following open string configuration in the
background (\ref{bac})
 \be
 t=\tau,\;\;\;\;\;\theta=\sigma,\;\;\;\;\; y=y(\theta),\;\;\;\;\;
 r=0\;.
 \ee
The corresponding classical action is given by
 \be
 S=\frac{T}{2\pi\alpha'}\int d\theta\sqrt{{y'}^2+\frac{y^2}{-{\tilde
 z}}}\;.
 \ee
To get an insight of what kind of physics one might get we shall consider the case where
$r_0\gg r_N$. Here $r_0$ is the turning point of string where $y'$ is zero. In this limit
one may expand ${\tilde z}$ for large $y$ getting
 \be
 S\sim \frac{T}{2\pi \alpha'}\int d\theta \sqrt{(\frac{dy}{d\theta})^2+y^2\frac{g_1}{g_0^2}
 (1+\frac{g_0}{g_1}y^2)}\;.
 \ee
Setting ${\tilde\theta}=\theta\frac{g_1^{1/2}}{g_0}$ one finds
 \be
 S\sim \frac{T}{2\pi \alpha'}\int d{\tilde \theta} \sqrt{(\frac{dy}{d{\tilde \theta}})^2+y^2
 (1+\frac{y^2}{R^2})}\;,
 \ee
where $R^2=g_1/g_0$. Using the result of the AdS space in the global coordinates one can now
easily read the potential of the external objects in terms of their distance in
$\theta$ direction.
In particular at leading order one gets
 \be
 E\sim -\frac{\sqrt{g_1}}{\theta}+ {\cal O}(\theta)\;,
 \ee
showing that the Wilson loop probes the background with the characteristic
length $\sqrt{g_1}$.


\section*{Conclusion}

In this paper we have studied the 1/2 BPS geometry of type IIB
which is characterized by concentric rings on the boundary
2-plane. The number of radii of the rings is taken to be odd such
that the background is asymptotically $AdS_5\times S^5$. Our probe
which explores this space time is a closed string that propagates
in the background according to the classical equations of motion.

The first configuration we have considered is the one studied in
\cite{Filev:2004yv}; a folded closed string extended in the radial
direction of $AdS$, spinning around its center of mass and
orbiting in $S^3$, and at the same time rotating in the $\tilde
{S}^3$ with a maximal radius. The one dimensional potential
governing the dynamics of these strings allows the radial
coordinate of the center of mass to be different from zero and
hence orbiting strings, in addition to spinning ones, appear in
the problem. This should be compared to $AdS_5\times S^5$ where
only spinning strings were allowed. As stated before, orbiting
configurations for strings are permitted in confining $AdS$
backgrounds and in this sense, the concentric rings show some
similarities with such backgrounds. In this work we extended the
results of \cite{Filev:2004yv} to cases with nonzero $\tilde
{S}^3$ angular momentum and found the energy dependence on the
angular momenta for short and long strings centered around the
origin for large values of the classical charges.

One could also consider folded configurations in the regions where
$\tilde {S}^3$ has shrunk to a point with the angular momentum
coming from rotations in the remaining angle of $S^5$. In the
$AdS_5\times S^5$ case, the string sigma model expansion around
such a configuration, in the point like string limit, would lead
to the pp wave background. In the present case one would
reasonably expect that a similar analysis could yield a similar
result. Our expectation is that for a shrunk $\tilde {S}^3$, the
new exterma are still present in the potential and these will
allow for several point like configurations such that the sigma
model expansion around each point would result in a pp wave like
limit.

Based on this expectation, we were able to find a limit of the
rings background which is a superposition of a number of pp waves.
For this purpose we considered the rings as a deformation of
$AdS_5\times S^5$ by taking very narrow rings at the edge of the
droplet. This superposition of pp waves is itself a solution of
the equations of motion which is characterized by a number of
horizontal black and white strips on the boundary plane bounded by
black and white regions from below and above respectively. We
called this background $\it{``Zebra"}$.

We also studied circular strings wound around the equator of $S^3$
and pulsating in the radial direction of $AdS$. The system reduces
to a one dimensional quantum mechanical system for this
configuration and we use Bohr-Sommerfeld analysis to find the
energy levels of the system when the level number is large. We
were able to do the calculations when the rings are considered as
a deformation of $AdS_5\times S^5$. This time the deformation is
produced by taking very narrow rings close to the center of the
droplet. The results show that unlike the folded strings,
pulsating ones experience no new physics in this background as
compared to $AdS_5\times S^5$. Our end result is an expansion for
the energy of the string in terms of the level (and winding)
number and the deformation parameters.

We have also studied Wilson loop of the corresponding dual theory,
using open strings on 1/2 BPS geometry (\ref{bac}) where we found
some corrections which could be due to short distance effects,
showing that the dual theory must be defined on a sphere with
characteristic length $\sqrt{g_1}$.

Finally we note that, although the results we have found are
qualitatively the same as those in $Ads_5\times S^5$, the strings
do see a new structure of the background. In fact as we have seen
different strings probe this background with different parameters.
For example in Wilson loop the parameter is given by $\sqrt{g_1}$,
while for the rotating short and long strings it is given by
$\sqrt{f_0}$ and $\sqrt{g_0}$, respectively.

\section*{Acknowledgments}
We would like to thank Mohsen Alishahiha for collaboration in the
early stage of this work and useful comments and discussions. We
would also like to thank Albion Lawrence and Soo-Jong Rey for
useful discussions.


\begin{thebibliography}{99}

\bibitem{Maldacena:1997re}
J.~M.~Maldacena,
``The large $N$ limit of superconformal field theories and supergravity,''
Adv.\ Theor.\ Math.\ Phys.\  {\bf 2}, 231 (1998)
[Int.\ J.\ Theor.\ Phys.\  {\bf 38}, 1113 (1999)]
[arXiv:hep-th/9711200].

\bibitem{Gubser:1998bc}
S.~S.~Gubser, I.~R.~Klebanov and A.~M.~Polyakov,
``Gauge theory correlators from non-critical string theory,''
Phys.\ Lett.\ B {\bf 428}, 105 (1998)
[arXiv:hep-th/9802109].



\bibitem{Witten:1998qj}
E.~Witten,
``Anti-de Sitter space and holography,''
Adv.\ Theor.\ Math.\ Phys.\  {\bf 2}, 253 (1998)
[arXiv:hep-th/9802150].

\bibitem{Sheikh-Jabbari:2004ik}
M.~M.~Sheikh-Jabbari,
``Tiny graviton matrix theory: DLCQ of IIB plane-wave string theory, a
conjecture,''
JHEP {\bf 0409}, 017 (2004)
[arXiv:hep-th/0406214].



\bibitem{Berenstein:2004kk}
D.~Berenstein,
``A toy model for the AdS/CFT correspondence,''
JHEP {\bf 0407}, 018 (2004)
[arXiv:hep-th/0403110].






\bibitem{Lin:2004nb}
H.~Lin, O.~Lunin and J.~Maldacena,
``Bubbling AdS space and 1/2 BPS geometries,''
JHEP {\bf 0410}, 025 (2004)
[arXiv:hep-th/0409174].





\bibitem{Buchel:2004mc}
A.~Buchel,
``Coarse-graining 1/2 BPS geometries of type IIB supergravity,''
arXiv:hep-th/0409271.


\bibitem{deMelloKoch:2004ws}
R.~de Mello Koch and R.~Gwyn,
``Giant graviton correlators from dual $SU(N)$ super Yang-Mills theory,''
arXiv:hep-th/0410236.

\bibitem{Gubser:2004xx}
S.~S.~Gubser and J.~J.~Heckman,
``Thermodynamics of R-charged black holes in $AdS_5$ from effective strings,''
arXiv:hep-th/0411001.

\bibitem{Suryanarayana:2004ig}
N.~V.~Suryanarayana,
``Half-BPS giants, free fermions and microstates of superstars,''
arXiv:hep-th/0411145.


\bibitem{Caldarelli:2004ty}
M.~M.~Caldarelli,
``On supersymmetric solutions of $D = 4$ gauged supergravity,''
arXiv:hep-th/0411153.


\bibitem{Filev:2004yv}
V.~Filev and C.~V.~Johnson,
``Operators with large quantum numbers, spinning strings, and giant
gravitons,''
arXiv:hep-th/0411023.

\bibitem{Gubser:2002tv}
S.~S.~Gubser, I.~R.~Klebanov and A.~M.~Polyakov,
``A semi-classical limit of the gauge/string correspondence,''
Nucl.\ Phys.\ B {\bf 636}, 99 (2002)
[arXiv:hep-th/0204051].

\bibitem{Frolov:2002av}
S.~Frolov and A.~A.~Tseytlin,
``Semiclassical quantization of rotating superstring in $AdS_5\times S^5$,''
JHEP {\bf 0206}, 007 (2002)
[arXiv:hep-th/0204226].

\bibitem{semi}
A.~A.~Tseytlin,
``Semiclassical quantization of superstrings: $AdS_5\times S^5$ and beyond,''
Int.\ J.\ Mod.\ Phys.\ A {\bf 18}, 981 (2003)
[arXiv:hep-th/0209116];

G.~Arutyunov, S.~Frolov, J.~Russo and A.~A.~Tseytlin,
``Spinning strings in $AdS_5\times S^5$ and integrable systems,''
Nucl.\ Phys.\ B {\bf 671}, 3 (2003)
[arXiv:hep-th/0307191];

G.~Arutyunov, J.~Russo and A.~A.~Tseytlin,
``Spinning strings in $AdS_5\times S^5$: New integrable system relations,''
Phys.\ Rev.\ D {\bf 69}, 086009 (2004)
[arXiv:hep-th/0311004];

S.~Frolov and A.~A.~Tseytlin,
``Multi-spin string solutions in $AdS_5\times S^5$,''
Nucl.\ Phys.\ B {\bf 668}, 77 (2003)
[arXiv:hep-th/0304255];

S.~Frolov and A.~A.~Tseytlin,
``Quantizing three-spin string solution in $AdS_5 \times S^5$,''
JHEP {\bf 0307}, 016 (2003)
[arXiv:hep-th/0306130];

S.~Frolov and A.~A.~Tseytlin,
``Rotating string solutions: AdS/CFT duality in non-supersymmetric  sectors,''
Phys.\ Lett.\ B {\bf 570}, 96 (2003)
[arXiv:hep-th/0306143];

M.~Alishahiha, A.~E.~Mosaffa and H.~Yavartanoo,
``Multi-spin string solutions in AdS black hole and confining backgrounds,''
Nucl.\ Phys.\ B {\bf 686}, 53 (2004)
[arXiv:hep-th/0402007].


\bibitem{Tseytlin:2004xa}
A.~A.~Tseytlin,
``Semiclassical strings and AdS/CFT,''
[arXiv:hep-th/0409296];

A.~A.~Tseytlin,
``Spinning strings and AdS/CFT duality,''
[arXiv:hep-th/0311139].



\bibitem{Armoni:2002xp}
A.~Armoni, J.~L.~F.~Barbon and A.~C.~Petkou,
``Orbiting strings in AdS black holes and ${\cal N} = 4$ SYM at finite  temperature,''
JHEP {\bf 0206}, 058 (2002)
[arXiv:hep-th/0205280].

\bibitem{Armoni:2002fr}
A.~Armoni, J.~L.~F.~Barbon and A.~C.~Petkou,
``Rotating strings in confining AdS/CFT backgrounds,''
JHEP {\bf 0210}, 069 (2002)
[arXiv:hep-th/0209224].


\bibitem{Minahan:2002rc}
J.~A.~Minahan,
``Circular semiclassical string solutions on $AdS_5\times S^5$,''
Nucl.\ Phys.\ B {\bf 648}, 203 (2003)
[arXiv:hep-th/0209047].

\bibitem{pulsating}
M.~Alishahiha and A.~E.~Mosaffa,
``Circular semiclassical string solutions on confining AdS/CFT backgrounds,''
JHEP {\bf 0210}, 060 (2002)
[arXiv:hep-th/0210122];

N.~Bobev, H.~Dimov and R.~C.~Rashkov,
``Pulsating strings in warped $AdS_6\times S^4$ geometry,''
[arXiv:hep-th/0410262];

M.~Smedback,
``Pulsating strings on $AdS_5\times S^5$,''
JHEP {\bf 0407}, 004 (2004)
[arXiv:hep-th/0405102];

H.~Dimov and R.~C.~Rashkov,
``Generalized pulsating strings,''
JHEP {\bf 0405}, 068 (2004)
[arXiv:hep-th/0404012];

A.~Khan and A.~L.~Larsen,
``Spinning pulsating string solitons in $AdS_5\times S^5$,''
Phys.\ Rev.\ D {\bf 69}, 026001 (2004)
[arXiv:hep-th/0310019].



\bibitem{Maldacena:1998im}
J.~M.~Maldacena,
``Wilson loops in large $N$ field theories,''
Phys.\ Rev.\ Lett.\  {\bf 80}, 4859 (1998)
[arXiv:hep-th/9803002].

\bibitem{Rey:1998ik}
S.~J.~Rey and J.~T.~Yee,
``Macroscopic strings as heavy quarks in large $N$ gauge theory and  anti-de
Sitter supergravity,''
Eur.\ Phys.\ J.\ C {\bf 22}, 379 (2001)
[arXiv:hep-th/9803001].

\bibitem{cusp}
N.~Drukker,~D.~J.~Gross and H.~Ooguri, ``Wilson loops and minimal
surfaces,'' Phys.\ Rev.\ D {\bf 60}, 125006 (1999)
[arXiv:hep-th/9904191].




\end{thebibliography}
\end{document}